\documentclass[fleqn,usenatbib,useAMS]{mnras}
\usepackage{amsmath}	% Advanced maths commands
\usepackage{amssymb}	% Extra maths symbols
\usepackage{multicol}        % Multi-column entries in tables
\usepackage{bm}		% Bold maths symbols, including upright Greek
\usepackage{graphicx,epsfig}
\usepackage{float}
\usepackage{xcolor}
\newcommand{\kms}{\,km\,\,s$^{-1}$} % kilometres per second
\newcommand{\PGPU}{$\varphi-$GPU }

\newcommand{\Mbh}{M_{\text{smbh}}}
\newcommand{\Mcl}{M_{\text{cl}}}
\newcommand{\Rcl}{R_{\text{cl}}}

\newcommand{\add}[1]{{\bf{#1}}}
\newcommand{\orcid}[1]{\href{https://orcid.org/#1}{\textcolor[HTML]{A6CE39}{\aiOrcid}}}
\usepackage[T1]{fontenc}
\usepackage{ae,aecompl}

\title[Counter-rotation in the torus of NGC~1068]{Apparent counter-rotation in the torus of NGC~1068: \\ influence of an asymmetric wind}
\author[Elena Yu. Bannikova,  et al.] {Elena Yu. Bannikova$^{1,2,3}$\thanks{Contact e-mail: \href{mailto:bannikova@astron.kharkov.ua}
{bannikova@astron.kharkov.ua}}, Nina O. Akerman$^{4,5}$  \author[0000-0001-7011-9291]{Nina O. Akerman}, Massimo Capaccioli$^{1,6}$, \and Peter P. Berczik$^{7,8,9}$, 
 Vladimir S. Akhmetov$^{10,3}$, Marina V. Ishchenko$^{8}$\\
$^{1}$ INAF - Astronomical Observatory of Capodimonte, Salita Moiariello 16, I-80131, Naples, Italy\\
$^{2}$ Institute of Radio Astronomy, National Academy of Sciences of Ukraine, Mystetstv 4, UA-61002 Kharkiv, Ukraine \\
$^{3}$ V.N.Karazin Kharkiv National University, Svobody Sq.4, UA-61022, Kharkiv, Ukraine \\
$^{4}$ Dipartimento di Fisica e Astronomia ``Galileo Galilei'', Universit{\`a} degli Studi di Padova, vicolo dell'Osservatorio 3, I-35122, Padova, Italy \\
$^{5}$ INAF - Astronomical Observatory of Padova, vicolo dell'Osservatorio 5, I-35122 Padova, Italy \\
$^{6}$ University of Naples ``Federico II'', C.U. Monte Sant'Angelo, via Cinthia, I-80126, Naples, Italy \\
$^{7}$ Astronomisches Rechen-Institut am Zentrum fuer Astronomie der Universitaet Heidelberg, 
Moenchhofstrasse 12-14, \\ D-69120 Heidelberg, Germany \\
$^{8}$ Main Astronomical Observatory, National Academy of Sciences of Ukraine, 27 Akademika 
Zabolotnoho St., UA-03143 Kyiv, Ukraine \\
$^{9}$ Konkoly Observatory, Research Centre for Astronomy and Earth Sciences, E\"otv\"os Lor\'and Research Network (ELKH), \\ MTA Centre of Excellence, Konkoly Thege Mikl\'os \'ut 15-17, 1121 Budapest, Hungary \\
$^{10}$ INAF - Osservatorio Astrofisico di Torino, via Osservatorio 20,
I-10025 Pino Torinese (TO), Italy} 

\date{Last updated **** ******* **; in original form **** ****** *}

\pubyear{2022}

\begin{document}
\label{firstpage}
\pagerange{\pageref{firstpage}--\pageref{lastpage}}
\maketitle

\begin{abstract}
The recent ALMA maps together with observations of H$_2$O maser emission seem to suggest the presence of a counter-rotation in the obscuring torus of NGC 1068. We propose to explain this phenomenon as due to the influence of a wind, considered as radiation pressure, and the effects of torus orientation. In order to test this idea: 1. we make $N$-body simulation of a clumpy torus  
taking into account mutual forces between particles (clouds); 2. we apply ray-tracing algorithm with the beams from the central engine to choose the clouds in the torus throat that can be under direct influence of the accretion disk emission; 3. we use semi-analytical model to simulate the influence of the asymmetrical radiation pressure (wind) forced on the clouds in the torus throat. An axis of such a wind is tilted with respect to the torus symmetry axis; 4. we orient the torus relative to an observer and again apply ray-tracing algorithm. In this step the beams go from an observer to the optically thick clouds that allows us to take into account the mutual obscuration of clouds; 5. after projecting on the picture plane, we impose a grid on the resulting cloud distribution and find the mean velocity of clouds in each cells to mimic the ALMA observational maps.  By choosing the parameters corresponding to NGC 1068 we obtain the model velocity maps that emulate the effect of an apparent counter-rotation and can explain the discovery made by ALMA.
\end{abstract}

\begin{keywords}
active galactic nuclei, Sy galaxy, NGC~1068, gravity.
\end{keywords}

%%%%%%%%%%%%%%%%% BODY OF PAPER %%%%%%%%%%%%%%%%%%

\section{Introduction}

The structure of active galactic nuclei (AGNs) is described by the unified scheme suggested in \citep{1985ApJ...297..621A} and further developed by \cite{1988ApJ...329..702K, 1993ARA&A..31..473A, 1995PASP..107..803U}. 
The main idea is that a toroidal structure surrounds a supermassive black hole (SMBH) with an accretion disk. This structure is known as obscuring torus because, being dusty and optically thick, it obscures the central engine. The unified scheme explains AGN types 1 and 2 by different orientation of the torus relative to the observer; see the sketch of the unified scheme for radio-quiet/radio-loud AGNs in \citep{2012agn..book.....B}. The obscuring torus is a reservoir of matter feeding an accretion disk and it plays an essential role in the AGN activity.  In this scheme the torus must be geometrically thick in order to reduce significantly the view of the central engine. Different mechanisms providing the substantial thickness of the torus  were suggested such as the elastic collisions of magnetized clouds \citep{1988ApJ...329..702K}, the support by infrared (IR) radiation pressure \citep{1992ApJ...399L..23P, Krolik2007}, by energy feedback from supernova explosions \citep{2002ApJ...566L..21W} 
or by the  circulation due to the presence of the inflow (accretion) and outflow (wind) as in the dipolar vortex model \citep{Bannikova2007} and radiation-driven fountain model \citep{Wada2012}. Outflows can also play a role in obscuration of the central engine and it was suggested as an alternative of the dusty torus: funnel-shaped wind with embedded clumps \citep{2000ApJ...545...63E}, radiation-hydrodynamic model of dusty wind supported by IR radiation pressure \citep{2011ApJ...741...29D, 2012ApJ...747....8D}. On the other hand, $N$-body simulations of the self-gravitating clumpy torus show that it keeps geometrical thickness due to more complicated cloud dynamics than in the case of continuum medium disk \citep{2012MNRAS.424..820B, 2021MNRAS.503.1459B}. In the question about continuum versus clumpy torus structure the answer is in favor of the cloud structure. Indeed, analyses of observational spectral energy distribution (SED) combined with that from radiation transfer modeling  support the clumpy model. Among these there are the models in 2D case \citep{Nenkova2008a, Nenkova2008b}, in 3D case \citep{2012MNRAS.420.2756S, 2017MNRAS.470.2578G, 2021ApJ...919..136N} with application to NGC 1068 \citep{2006A&A...452..459H}, to Circinus \citep{Schartmann2008, 2017MNRAS.472.3854S}, and the disk+wind model: the hydrodynamic approach \citep{2020ApJ...897...26W} or the static distribution of clumps by some law \citep{2017ApJ...838L..20H, 2017MNRAS.472.3854S, 2019MNRAS.484.3334S}. 
All these components (accretion disk, torus, winds, clumps) and corresponding physical mechanisms are in play and all of them can be the reason of the complicated dynamics in AGNs.

The unified scheme was first applied to the Seyfert type 2 galaxy NGC~1068 which became a cornerstone object for model verification. Resent observations showed that NGC~1068 has an unusual dynamics which required an increase in the model complexity. We will shortly remind the main observations related to understanding the dynamics in this AGN. VLA and VLBI observations of water masers in its central region revealed complicated dynamics with non-keplerian motion and with an inclined maser disk relative to the equatorial plane \citep{Greenhill1996, Gallimore1996}. 
There is also asymmetry in distribution of the red and blue masers. 
Moreover, the maser spots have a velocity gradient, which confirms the clumpy nature of the torus. 
The maser emission coincides with the radio-continuum source S1 which is identified as the central engine \citep{1996ApJ...458..136G, 2004ApJ...613..794G}.

Direct observations of the torus in NGC 1068 in the IR band made with VLTI/MIDI disclosed a two-component structure in the temperature distribution that is related to the emission of the throat (hot) and of the main body (warm) of the torus \citep{Jaffe2004, Raban2009}. 
\cite{2007A&A...474..837T} found a similar result with the temperature stratification in the torus of Circinus galaxy where  water maser emission also presents \citep{Greenhill2003}. Interferometric observations of Seyfert AGNs support a scenario where mid-infrared (mid-IR) emission originates in the polar region from a biconic dusty wind launched in the inner hot part of the torus
\citep{2012ApJ...755..149H, 2013ApJ...771...87H, 2014A&A...563A..82T}. 

ALMA made the next important step in understanding the dynamics in the torus of NGC~1068. \cite{2016ApJ...823L..12G} resolved the dusty torus; they found a minor-to-major axis ratio $0.8 \pm 0.1$, and estimated the mass of the molecular gas to be $(1\pm 0.3) \times 10^5 M_\odot$ and the radius $R_{\text{torus}} = 3.5 \pm 0.5$~pc. 
Their maps show a velocity gradient along the minor axis of the torus. They also reported, by H$_2$O maser emission, an apparent counter-rotation of the outer disk with a $3.5$~pc radius relative to the inner $1$~pc region. ALMA HCN(3-2) and HCO$^+$ (3-2) maps of the torus show similar results \citep{2018ApJ...853L..25I}. \cite{Impellizzeri2019} presented 1.4~pc (20 mas) resolution observations of 256~GHz radio continuum and HCN (3-2) maps of the torus in NGC~1068. Their radio observations resolved an X-shaped structure with size comparable to that of the torus. This X-shape emission feature can be a region where the gas in the torus throat interacts with the wind.
They also discovered that the inner disk (inside 1.2~pc) coincides with the H$_2$O megamaser disk, and it counter-rotates relative to the outer disk that extends to $7$~pc.

The next ALMA observations were made by \cite{2019A&A...632A..61G}. They mapped the emission of the molecular gas tracers in the circumnuclear disk (CND) and in the torus (the CO(2-1), CO(3-2), and HCO$^+$ (4-3) lines), as well as continuum emission with high spatial resolution up to $0.03''\approx 2$~pc. They discovered an elongated molecular torus with mass $M_{\text{torus}} = 3 \times 10^5 M_\odot$ and size $11 \pm 0.6$~pc in the HCO$^+$ line (in the CO lines it is bigger). They also confirmed counter-rotation discovered in the previous observations. %\cite{2019A&A...632A..61G} argued that such 
A configuration of two counter-rotating disks is dynamically unstable and will collapse in a timescale of one orbital period. 
Indeed, the counter-rotation of two layers will lead to the Kelvin-Helmholtz (KH) instability between the reversely rotating parts and, as a result, to  unavoidable catastrophe. \cite{2020MNRAS.497.1020W} showed that the close binary SMBH provides tidal torques to prevent the torus from the KH catastrophe but this idea requires the additional proof of presence a binary SMBH and to taking into account the clumpy structure in the torus. \cite{2019A&A...632A..61G} suggested that the apparent counter-rotation can be the result of the projection of the outflows that are present in the torus due to the interaction with the AGN wind. The wind opening angle of $\theta =80^\circ$ satisfies the kinematics in the torus. \cite{2020ApJ...897...26W} made hydrodynamical simulations and found some effects of the apparent counter-rotation which appears due to the presence of wind. 

In this paper we use our dynamical model of clumpy torus
 adding the outflow forces in order to check whether the apparent counter-rotation in the torus in NGC~1068 can be explained by the influence of the wind related to the radiation from accretion disk. In Section~\ref{sec:model} we present $N$-body simulation of a self-gravitating clumpy torus and the semi-analytical model of the asymmetrical wind related to the radiation from the accretion disk. 
In Section~\ref{sec:contrrot}  we  build velocity maps for the torus taking into account the mutual obscuration of clouds and the effects of the torus orientation. We summarize our conclusions in Section~\ref{sec:conclusions}.

\section{Model}
\label{sec:model}
In this paper  we use our model \citep{2012MNRAS.424..820B, 2021MNRAS.503.1459B} taking into account the self-gravity of the torus, the influence of the radiation pressure force, the orientation of the torus relative to an observer, and the mutual obscuration of the clouds. The final result are the velocity maps. 
The simulations consist on the following steps.

\begin{itemize}
\item {\it N-body simulation} of the self-gravitating clumpy torus in the gravitational field of a SMBH in the collisionless approach. 
Since the distance from the SMBH to the nearest particle is larger than the accretion disk radius, we are treating SMBH as the classical point mass. We also treat the particles as clouds since we are using Plummer potential with the softening parameter that play a role of the cloud scale. 
We ignore the possible changes of the cloud shapes during its interactions and all external influence (feedback of supernova explosions, external accretion, etc.). 

\item {\it Acting of the wind force.}  
We consider the wind force as radiation pressure from the accretion disk acting on the clouds in the torus throat. We apply ray-tracing algorithm with the beams from the central engine to choose these clouds. We do not include the wind as a form of additional force in our $N$-body simulations, but estimate its influence based on a semi-analytic model. We can use such approach because the action time by the wind force on the cloud is negligibly smaller than the orbital period of a torus. 

\item {\it Effects of obscuration}. We consider the clouds in the torus as optically thick. In this case, the clouds obscure each other, and this affects the final velocity distribution seen by an observer depending on different torus orientations. Here we use a ray-tracing algorithm to extract these clouds and to use them in the following step.

\item {\it Velocity maps}. We project the final distribution of clouds on the picture plane and divide it into cells. To reconstruct a velocity map, we find the mean velocity in each cell. 
\end{itemize}

\subsection{The structure of the torus and its throat from $N$-body simulation}
\label{sec:nbody_throat}

\begin{figure}
\centering
\includegraphics[width = 85mm]{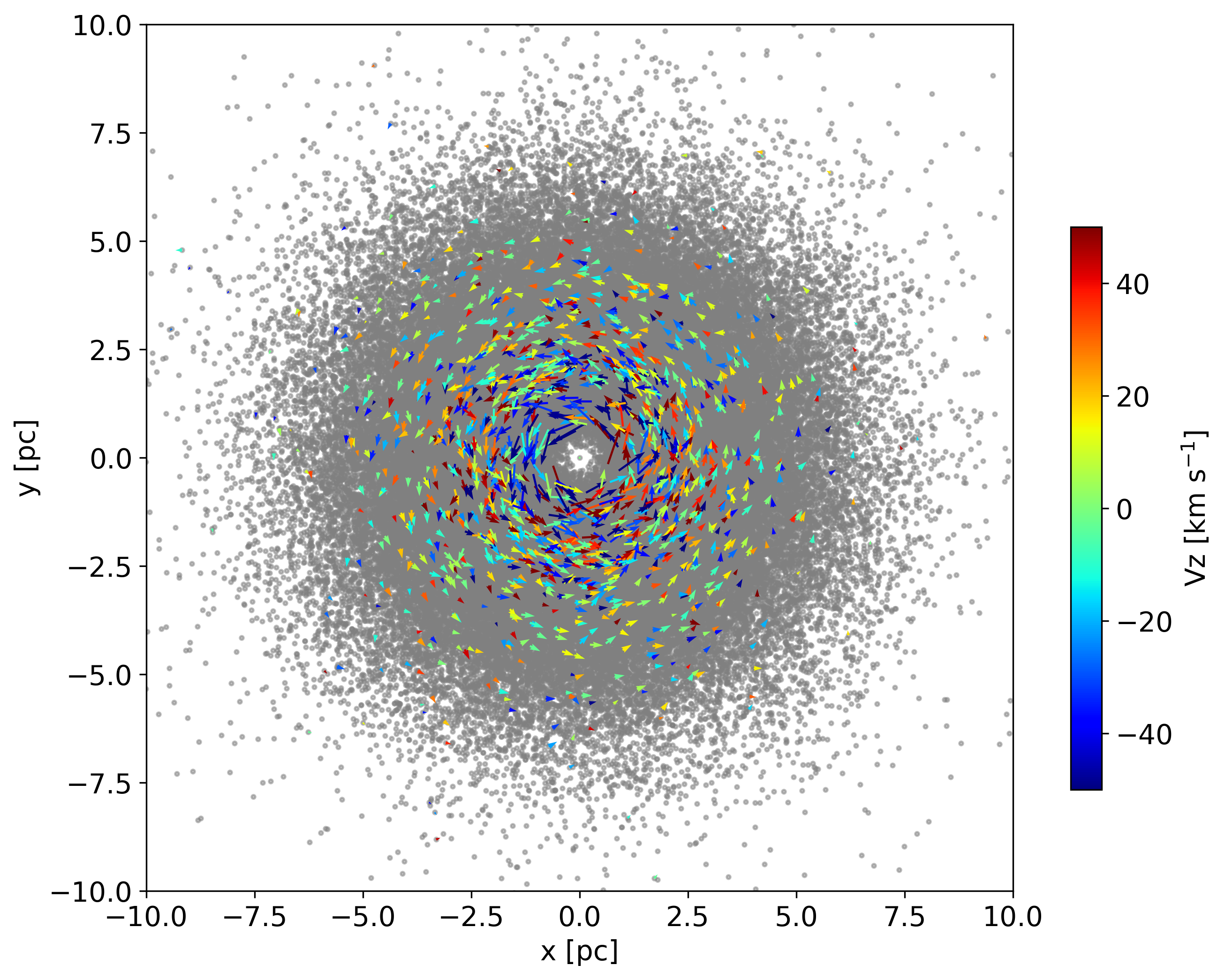}
\includegraphics[width = 85mm]{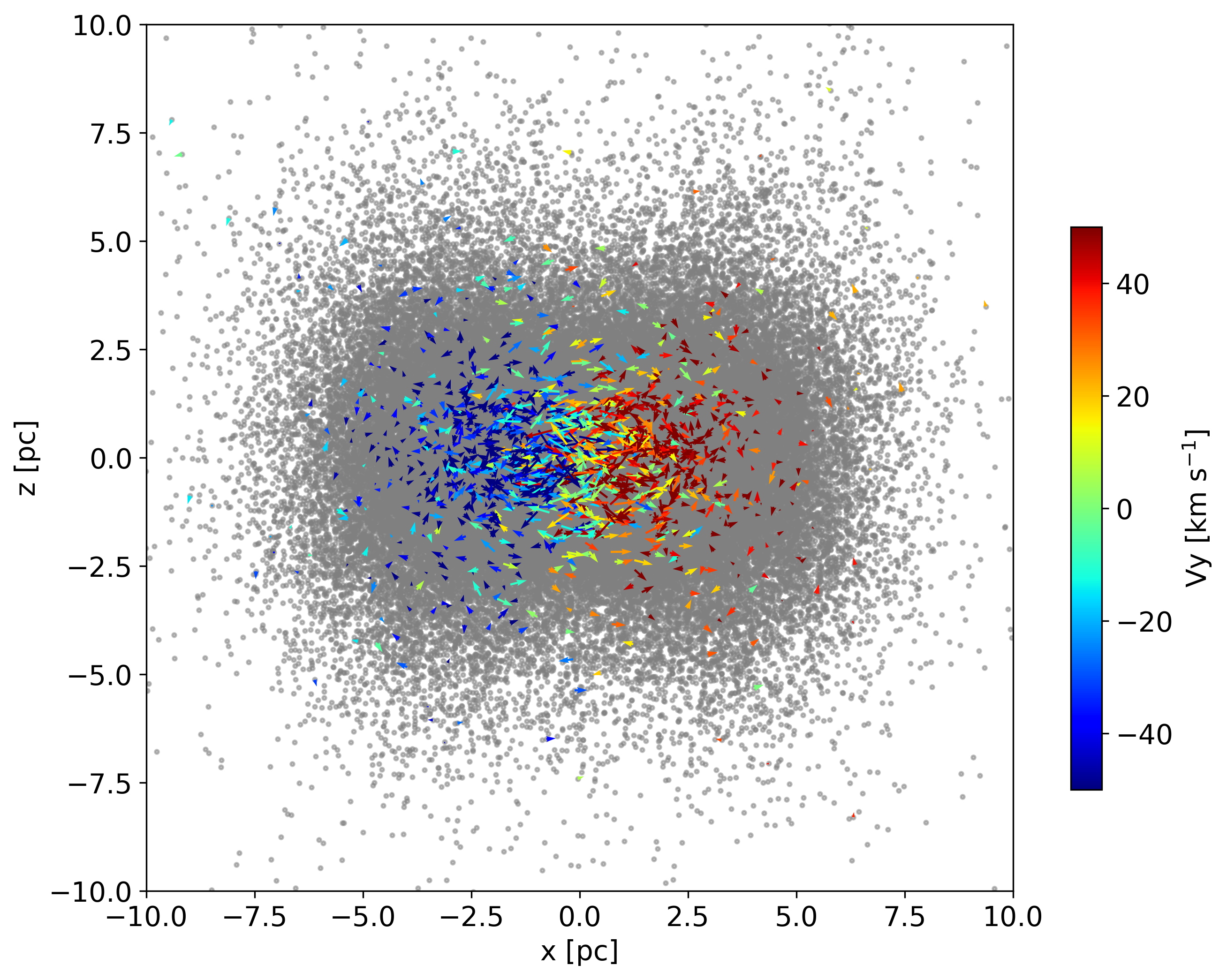}
\caption{Distribution of clouds for the torus in the state of equilibrium: face-on ({\it top}), edge-on ({\it bottom}). The coloured arrows show the direction and value of the corresponding velocity components.}
\label{fig:arrows-field}
\end{figure} 

\begin{figure}
\centering 
\includegraphics[width = 80mm]{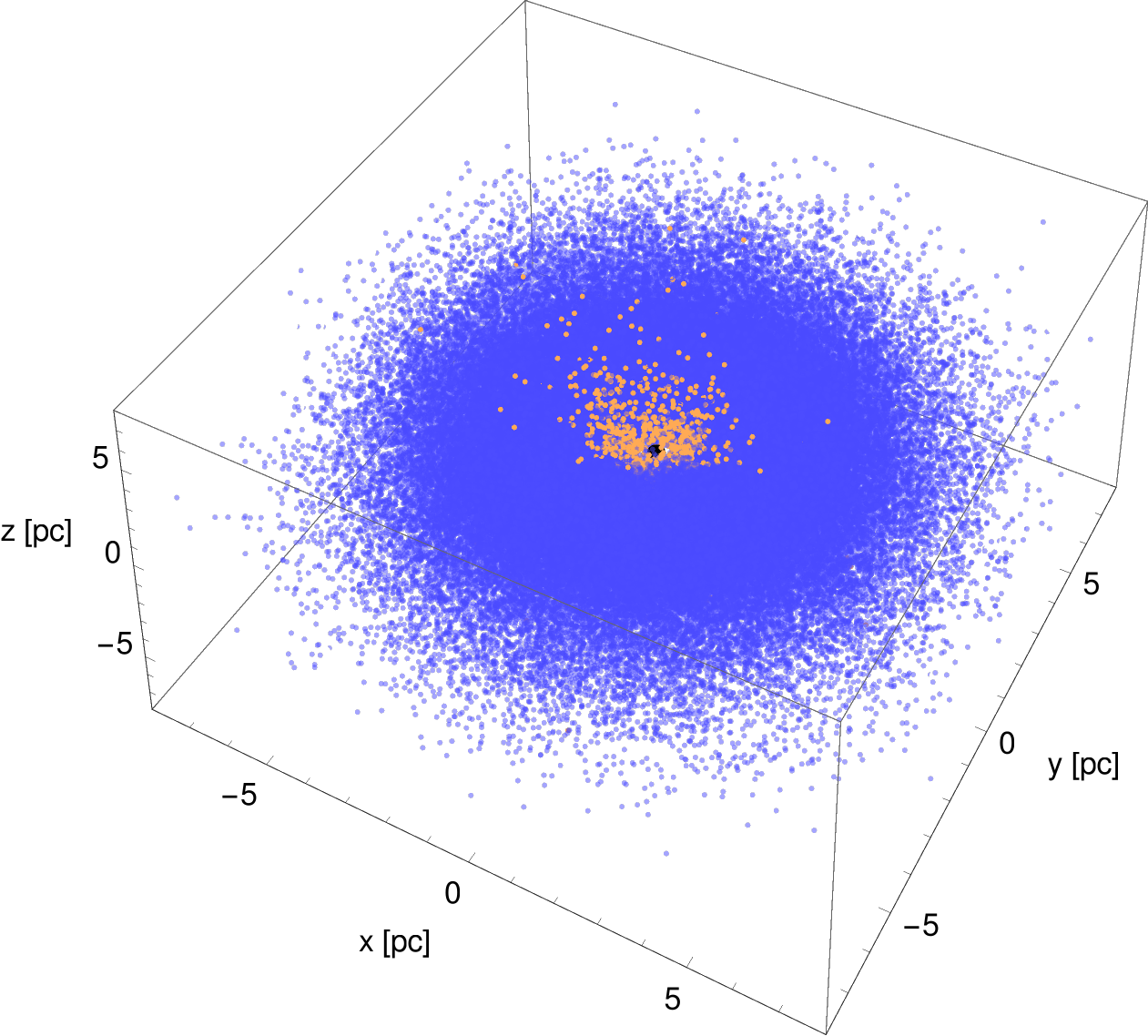}
\caption{Distribution of $N=10^5$ clouds of a torus in the equilibrium state ({\it blue}). Clouds in the torus throat ({\it yellow}) are obtained using the ray-tracing procedure from the central engine to each cloud.
The black point marks the position of SMBH.}
\label{fig:AllClouds3D}
\end{figure}

\begin{figure}
\centering 
\includegraphics[width = 80mm]{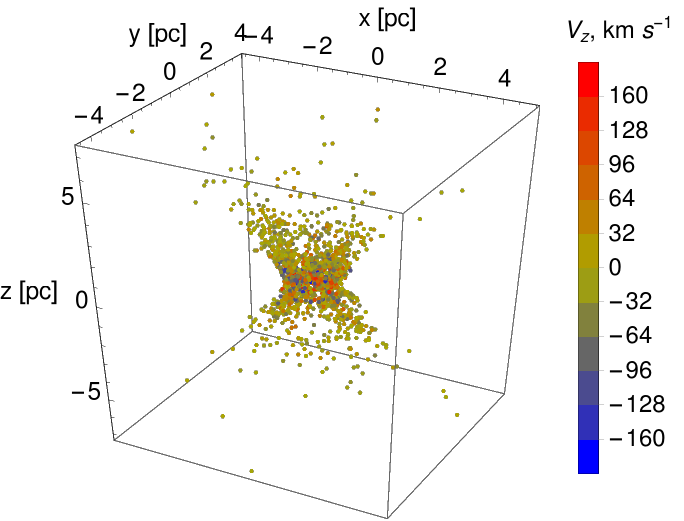}
\caption{Cloud distribution in the torus throat colour-coded by the velocity component $v_z$. The dimensionless cloud radius is $\varepsilon_\text{cl} =0.025$.}
\label{fig:throat3D}
\end{figure}

As the first step of our investigation we consider $N$-body problem of the massive particles moving around the SMBH taking into account the gravitational interactions among them. This allows us to obtain the equilibrium distribution of clouds in the torus due to self-gravity.
We solve numerically the equations of motion for $N$ clouds in the gravitational field of a central mass ($M_\text{smbh}$):
\begin{equation}\label{eq2.1}
  \mathbf{a}_k = -\frac{GM_\text{smbh}}{R_\text{torus}^2}\left[
  \frac{\mathbf{r}_k}{r_k^3} + \sum_{j=1}^N m_j \frac{\mathbf{r}_k - \mathbf{r}_j}{\left(|\mathbf{r}_k - \mathbf{r}_j|^2 + \varepsilon^2
\right)^{3/2}}
\right],
\end{equation}
where ${\mathbf{a}}_k$ is the acceleration of the $k$-th cloud and \mbox{$\mathbf{r}_k = (x_k, y_k, z_k)$} is its radius-vector normalized to the mean major radius of the torus $R_\text{torus}$. The relative mass $m_j$ of the particle (cloud) is normalized to the SMBH mass $M_\text{smbh}$ and the softening parameter  $\varepsilon$ corresponds to a dimensionless radius of cloud. One of the parameters is the relative mass of the torus, which we choose $M_\text{torus}=0.02 M_\text{smbh}$. This value corresponds to the mass range estimated from observations. Indeed, for the SMBH mass in NGC 1068 $M_\text{smbh} = (0.9 - 10) \times 10^6 M_\odot$ \citep{2018ApJ...853L..25I, Impellizzeri2019, 2021MNRAS.503.1459B} and the torus mass $M_\text{torus} = 3 \times 10^5 M_\odot$ \citep{2019A&A...632A..61G}. In this case, the relative torus mass is from a few percent to a few tens of percent of the SMBH mass. Our simulations of a self-gravitating torus show that the cloud distribution in the torus cross-section does not change significantly for a torus-SMBH mass ratio up to 0.1 (Bannikova et al., {\it in preparation}). We choose the total number of clouds $N=10^5$ that satisfies the obscuration condition \citep{Nenkova2008a, Nenkova2008b}. 

To choose the initial conditions, we use the following scenario of the torus formation \citep{2017FrASS...4...60B}. We can suggest that at the pre-active stage the clouds move around SMBH with random distribution in all orbital elements. The beginning of the active stage is related to the increase of the accretion rate and to the appearing of the radiation pressure forces. Since radiation pressure acts against the force of gravity, the clouds are pushing in two opposite directions, creating outflow cones. The clouds which are outside these cones continue to move around the SMBH. The final shape of the toroidal structure is formed due to self-gravity and it can be obtained using $N$-body simulation. We realize this idea in a simplified way using initial condition with  
random distribution of clouds in all orbital elements, forcing two hollow cones in two opposite directions along $z$-axis -- see Fig. 1 in \citep{2021MNRAS.503.1459B}. This corresponds to the simple representation of the beginning of active stage of AGN. The half-opening angle of the cones is $\theta =45^\circ$; it fits with the recent discussion in \citep{2019A&A...632A..61G, 2006AJ....132..620D}. 

The equations of motion (\ref{eq2.1})  were solved numerically using a high order Hermite integration scheme with the help of the \PGPU{\footnote{\url{https://github.com/berczik/phi-GPU-mole}}} code \citep{Wang2014, Zhong2014, Li2012, Li2017}. This code is fully parallelised and using the native GPU support with the NVIDIA CUDA library and it was used for many others astrophysical problems  \citep{Panamarev2018, 2019MNRAS.484.3279P, 2012ApJ...758...51J, 2018ApJ...868...97K}. 
In our previous paper \citep[see Section 3,][]{2021MNRAS.503.1459B} we run a separate study of the numerical $N$-body system resolution (particle numbers) and time integration (energy conservation). For the production runs in the current paper we use the largest particle number $N=10^5$ and the numerical integration parameter \citep{Aarseth1985} $\eta = 0.01$. With such parameters the global energy drift $(dE_{tot}/dt)$ of the system was at a level of $\sim 10^{-12}$. Our largest $N=10^5$ particle simulation up to the 1000 rotational period of the torus ($t_{end}$ = 6400) on our GPU computing system (with NVIDIA K20 GPU) requires $\sim$ 140 hour of real computation time with the total code performance around $\sim$1.15 Tflops.

As a result of this numerical simulations we obtain coordinates and velocity components of the clouds for the final equilibrium state of the torus.  Fig.~\ref{fig:arrows-field} shows the distribution of the vertical component of velocities for some clouds in two projections; it shows complicated dynamics which differs significantly from the simple Keplerian one. The values of the velocity were calculated here and in all the figures below for the following physical parameters of NGC~1068: $M_\text{smbh} = 5 \times 10^6 M_\odot$  and a mean major radius of the torus $R_\text{torus} = 3.5$~pc. We use this estimation of SMBH mass in NGC 1068 obtained  in  \citep{2020A&AT...31..423B, 2021MNRAS.503.1459B}.

Since we are interested in the dynamics of the clouds under the influence of outflows from the accretion disk,  we need to properly select such clouds. They are located in the torus throat near the bicones. To select them we use a ray-tracing algorithm: if a ray from the central engine (SMBH with accretion disk) meets a cloud, we extract and use this cloud in the following analysis. The main parameter in this case is the radius of a cloud 
$R_\text{cl} = \varepsilon_\text{cl} R_\text{torus}$, where 
$\varepsilon_\text{cl}$ 
coincides with a softening parameter in the N-body simulations. We showed in our previous papers that small changes of the softening parameter do not alter the equilibrium cross-section of the torus. On the other hand, the softening parameter defined as a relative size of a cloud plays a role in the effects of obscuration. Since we assume each cloud as optically thick, they obscure each other, which can influence the resulting velocity maps. In the following, we will consider two values of $\varepsilon_\text{cl} = 0.01$ and 0.025. They were used in our previous papers to satisfy the ALMA velocity and velocity dispersion maps.

Fig.~\ref{fig:AllClouds3D} shows the 3D distribution of the clouds in the equilibrium state of the torus (blue) with the central throat (yellow) obtained from the ray tracing algorithm for $\varepsilon_\text{cl} = 0.025$.  It means that these clouds in the torus throat will be under direct influence of the wind and experience an acceleration due to the radiation pressure.  The distribution of $v_z$ velocity component without the influence of the wind is symmetrical with respect to $z$-axis (Fig.~\ref{fig:throat3D}).
The velocity component is not high enough to explain the observed velocity of the maser spots, which further emphasizes the need for additional acceleration from the wind. To test our idea to explain the asymmetry of maser distribution and the counter-rotation, we will use a model of asymmetrical wind which we describe in the next subsection.

\subsection{The structure of the wind and its influence on the clouds in the torus throat} 
\label{sec:wind}

\begin{figure*}
\centering
\includegraphics[width = 140mm]{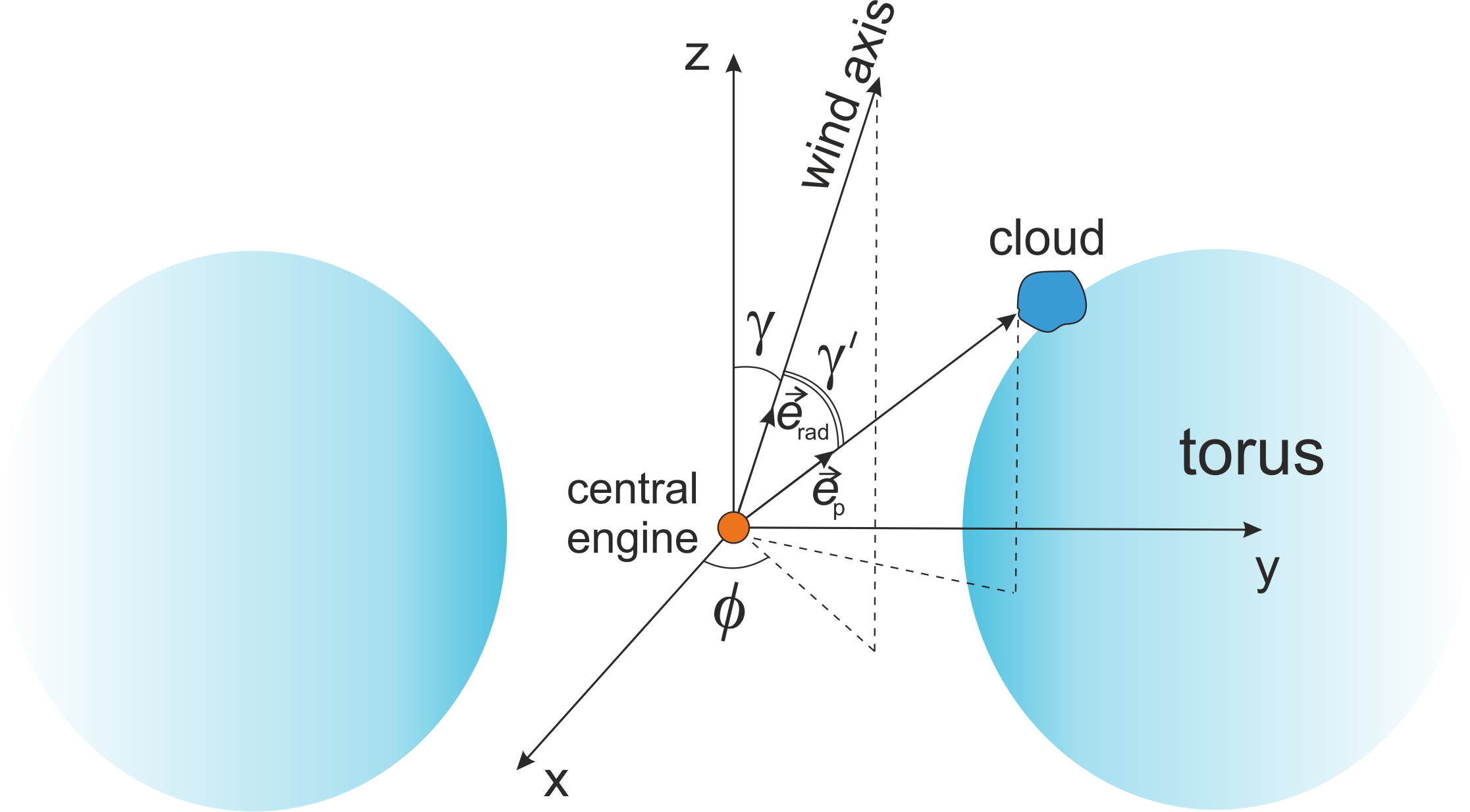}
\caption{Scheme of the wind orientation relative to the torus.}
\label{fig:windscheme}
\end{figure*}
We consider the wind as radiation pressure from the accretion disk. Radiation pressure force acting on a cloud is:
\begin{equation}\label{eqw1}
{\bf F}_{\text{rad}} = \frac{\sigma}{c}{\it F (r)} \, {\bf e}_{\text{rad}},
\end{equation}
where $\sigma$ is the cloud cross-section, $c$ the speed of light, and
${\bf e}_{\text{rad}}$  the unit vector along the wind axis.
The radiation flux from an isotropic point source with luminosity $L$ is:
\begin{equation}\label{eqw2}
F(r) = \frac{L}{2\pi r^2} .
\end{equation}
For convenience, we will use the relative luminosity $l=L/L_{\text{edd}}$, where $L_{\text{edd}}$ is the Eddington luminosity. 
Taking into account the cloud cross-section $\sigma = \pi \Rcl^2 = \pi \varepsilon_\text{cl}^2 R_{\text{torus}}^2$, 
we have:
\begin{equation}\label{eqw3}
|{\bf F}_{\text{rad}}| = \frac{\pi G m_{p}}{\sigma_{\text{T}}} \, \frac{\Mbh R_{\text{torus}}^2}{r^2}\, l \varepsilon_\text{cl}^2.
\end{equation}
The ratio of the wind force to the gravitational force\footnote{This expression (\ref{eqw4}) matches that obtained by \cite{2020ApJ...900..174V} if we consider $\Rcl = N_H/(2n_H)$. In this case, the column density of the obscuring structure corresponds to a cloud. In our consideration, the column density consists of the number of clouds on the line of sight.} is:
\begin{equation}\label{eqw4}
k=\frac{|{\bf F}_{\text{rad}}|}{|{\bf F}_{\text{gr}}|} = \frac{\pi m_{p}}{\sigma_{\text{T}}\Mcl}\, R_{\text{torus}}^2 \, l \varepsilon_\text{cl}^2.
\end{equation}
Using $\Mcl \approx M_\odot$, $\varepsilon_\text{cl} =0.025$ \citep{2021MNRAS.503.1459B},  \mbox{$R_{\text{torus}} = 3.5$}~pc,  
and $l \approx 0.1$, we have $k \approx 30$. So, in our simulations we will consider $1<k<30$.
\begin{figure}
\centering
\includegraphics[width = 80mm]{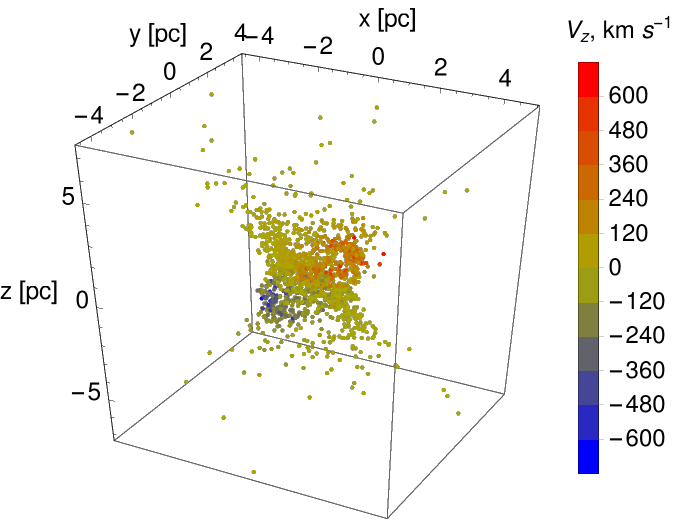}  
\includegraphics[width = 80mm]{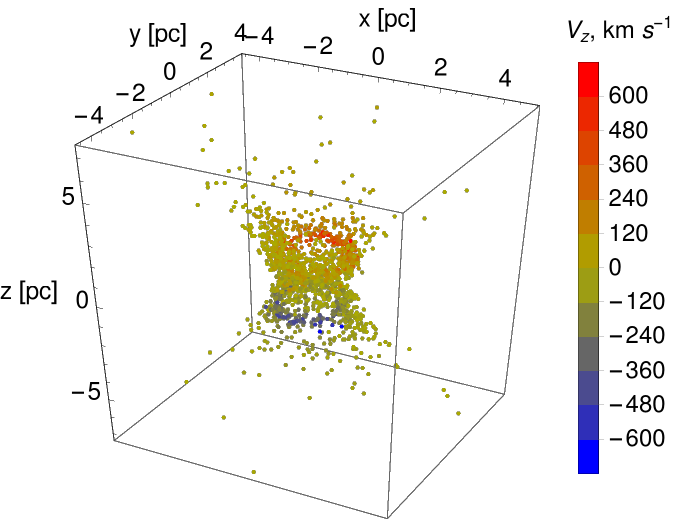}
\caption{Same as Fig.~\ref{fig:throat3D} after the wind has acted for a some short time interval. The wind parameters are: $k=20$, $\gamma=30^\circ$; $\phi=0^\circ$ ({\it top}), $\phi=90^\circ$ ({\it bottom}).}
\label{fig:hollownwind3D}
\end{figure} 

In order to test the idea that an asymmetrical wind can be responsible for the apparent counter-rotation in the torus, we tilt the wind axis with respect to the torus symmetry axis. The scheme of the wind orientation is shown in  Fig.~\ref{fig:windscheme}. We can represent the vector of the wind force acting on the $i$-th cloud as: 
\begin{equation}\label{eqw5}
{\bf F}_{\text{wind}} = k \cos\gamma' \frac{{\bf r}_i}{r_i^3} ,
\end{equation}
where $\gamma'$ is an angle between the wind axis and the direction to the cloud. So, we have:
\begin{equation}\label{eqw6}
\cos\gamma' =({\bf e}_{\text{p}} \cdot {\bf e}_{\text{rad}}).
\end{equation}
The unit vector along the wind axis in the reference system $(x,y,z)$ of the torus has the following components:
\begin{equation}\label{eqw7}
{\bf e}_{\text{rad}} =(\sin\gamma \cos\phi, \sin\gamma \sin\phi, \cos\gamma),
\end{equation}
where $\phi$ is the azimutal angle and $\gamma$ is the polar angle between the symmetry axis of the torus ($z$) and the wind axis.
The unit vector along the direction to the cloud is:
\begin{equation}\label{eqw8}
{\bf e}_{\text{p}} =\left(\frac{x}{r}, \frac{y}{r}, \frac{z}{r} \right),
\end{equation}
where $r=\sqrt{x^2 +y^2 +z^2}$.
So, we have:
\begin{equation}\label{eqw9}
\cos\gamma' = \frac{1}{r}(x\sin\gamma \cos\phi + y \sin\gamma \sin\phi + z \cos\gamma).
\end{equation}
Under the influence of both the gravitational force of SMBH and the wind force, the acceleration of $i$-th cloud is found by solving the equation of motion. In dimensionless units ($G=1$, $M_\text{smbh}=1$, \add{$R_\text{torus}=1$}):
\begin{equation}\label{eqw11}
{\bf a}_i = -\frac{{\bf r}_i}{r^3_i} + {\bf F}_{\text{wind}}, 
\end{equation}
where ${\bf r}_i$ is the dimensionless radius-vector of a cloud in torus throat. As initial condition we use the values of ${\bf r}_i$ and the correspondent velocities obtained by $N$-body simulation together with ray-tracing procedure (Fig.\ref{fig:AllClouds3D}, yellow).
This approach suggests that on considered stage the mutual gravitational forces exerted among clouds in the torus throat are negligibly small compared to the forces by SMBH and wind. 

We solve eq.~(\ref{eqw11}) to find coordinates and velocities that the clouds in the torus throat get under the influence of wind. We apply the wind force during a time interval of $\bigtriangleup t$=0.05 unit orbital period that is short enough to allow the clouds to stay in the torus throat. Finally, we transform the dimensionless coordinates and velocities to the dimensional ones for the SMBH mass and torus mean radius chosen above.

Fig.~\ref{fig:hollownwind3D} demonstrates structure and velocity distribution of the clouds in the torus throat after their acceleration by the asymmetrical wind, where the wind parameters are the polar angle $\gamma$ and the azimuthal angle $\phi$.  Here, the $y$ axis is the direction to the observer. We can see that the velocity distribution is asymmetrical along $z$ axis for $\phi = 0$ (Fig.~\ref{fig:hollownwind3D} {\it top})  and is symmetrical for $\phi = 90^\circ$ (Fig.~\ref{fig:hollownwind3D} {\it bottom}). The velocity distribution becomes asymmetric when the projection of the wind axis on the picture plane ($xz$ here) does not coincide with the symmetry axis of the torus. The asymmetry in 3D velocity distribution should influence the resulting velocity maps that we consider in the next Section.

\section{Results: velocity maps and apparent counter-rotating effect in the torus of NGC~1068}
\label{sec:contrrot}

\begin{figure*}
\centering
\includegraphics[width = 120mm]{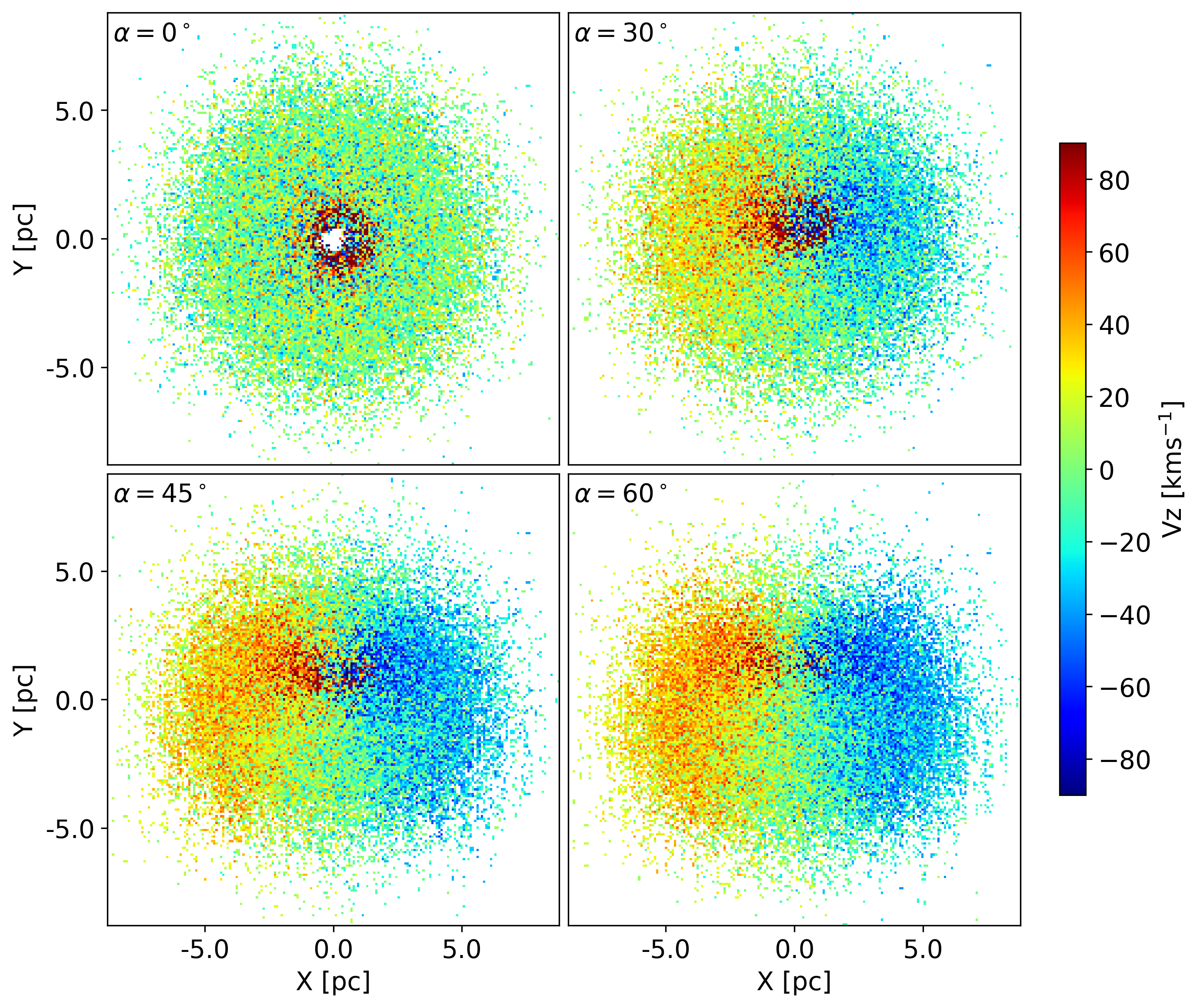}
\caption{Velocity maps in the picture plane $XY$ upon the influence of the wind for different orientations of the torus relative to the observer: $\alpha = 0^\circ, 30^\circ, 45^\circ, 60^\circ$.  The relative size of a cloud is $\varepsilon_\text{cl}=0.025$ and the wind azimuthal angle $\phi=0^\circ$.}
\label{fig:XY0}
\end{figure*} 

\begin{figure*}
\centering
\includegraphics[width = 120mm]{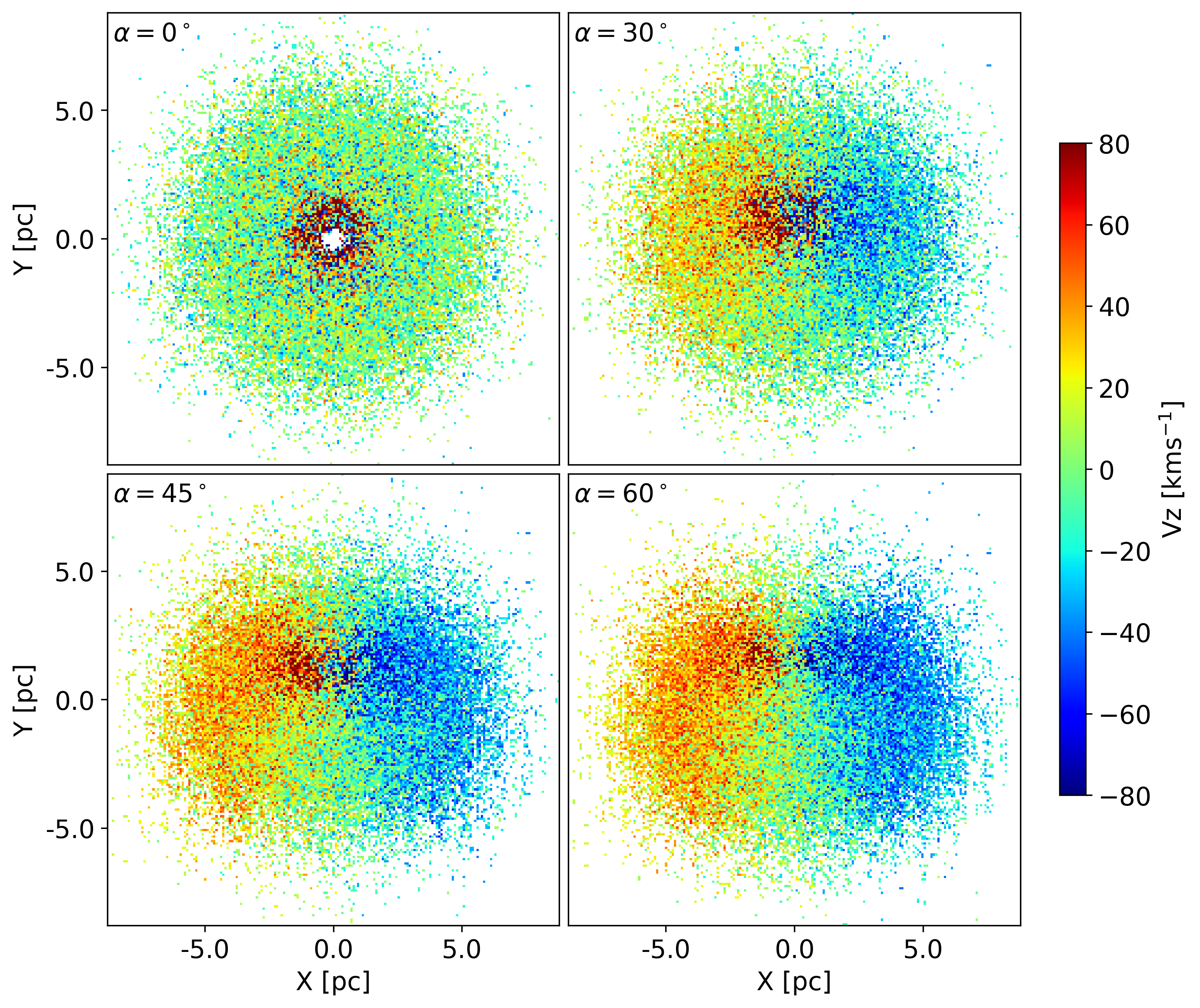}
\caption{Same as Fig.~\ref{fig:XY0} but $\phi=90^\circ$.}
\label{fig:XY90}
\end{figure*} 

\begin{figure*}
\centering
\includegraphics[width = 120mm]{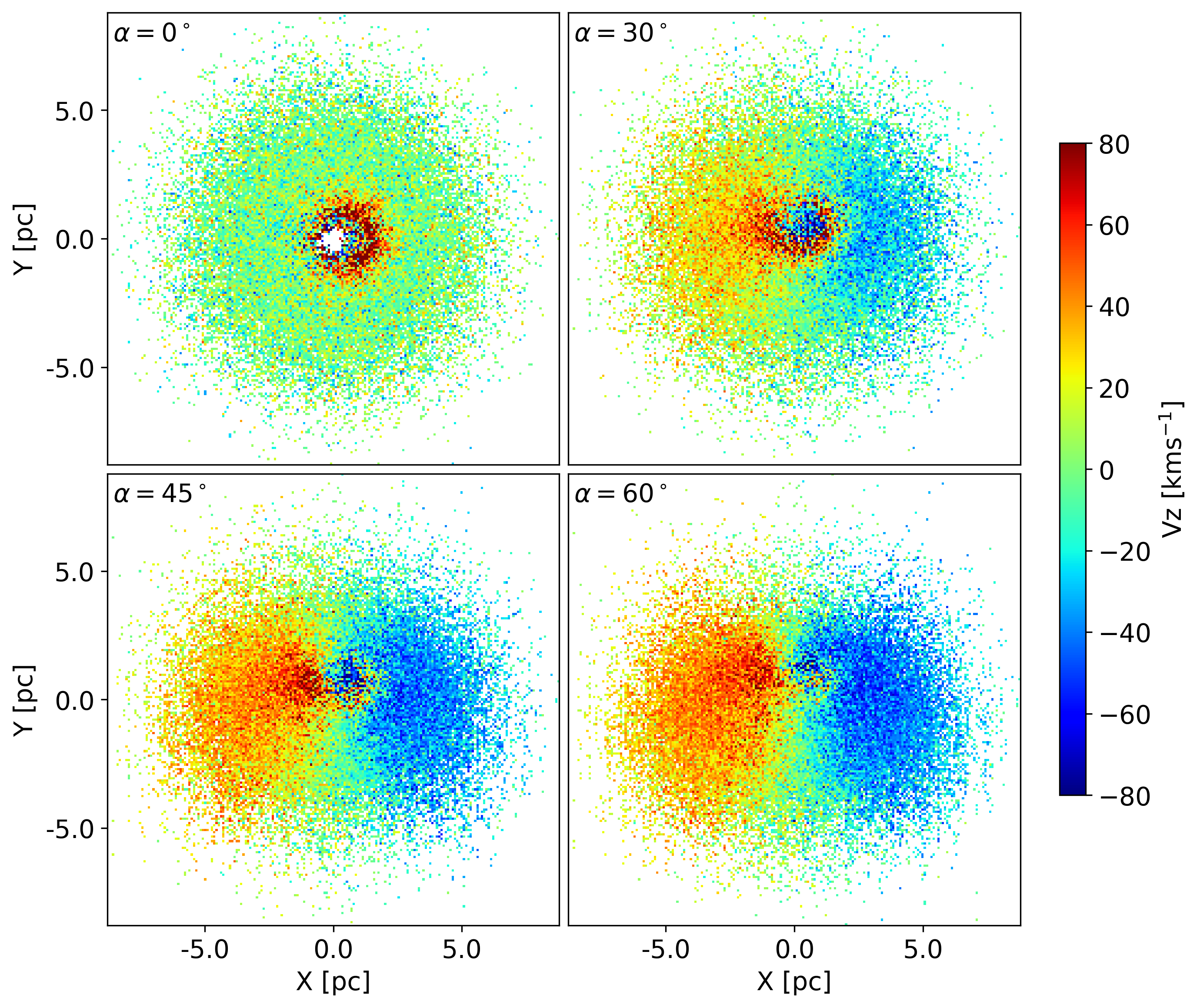}
\caption{Same as Fig.~\ref{fig:XY0} but $\varepsilon = 0.01$.}
\label{fig:XY0-eps0.01}
\end{figure*}

\begin{figure*}
\centering
\includegraphics[width = 120mm]{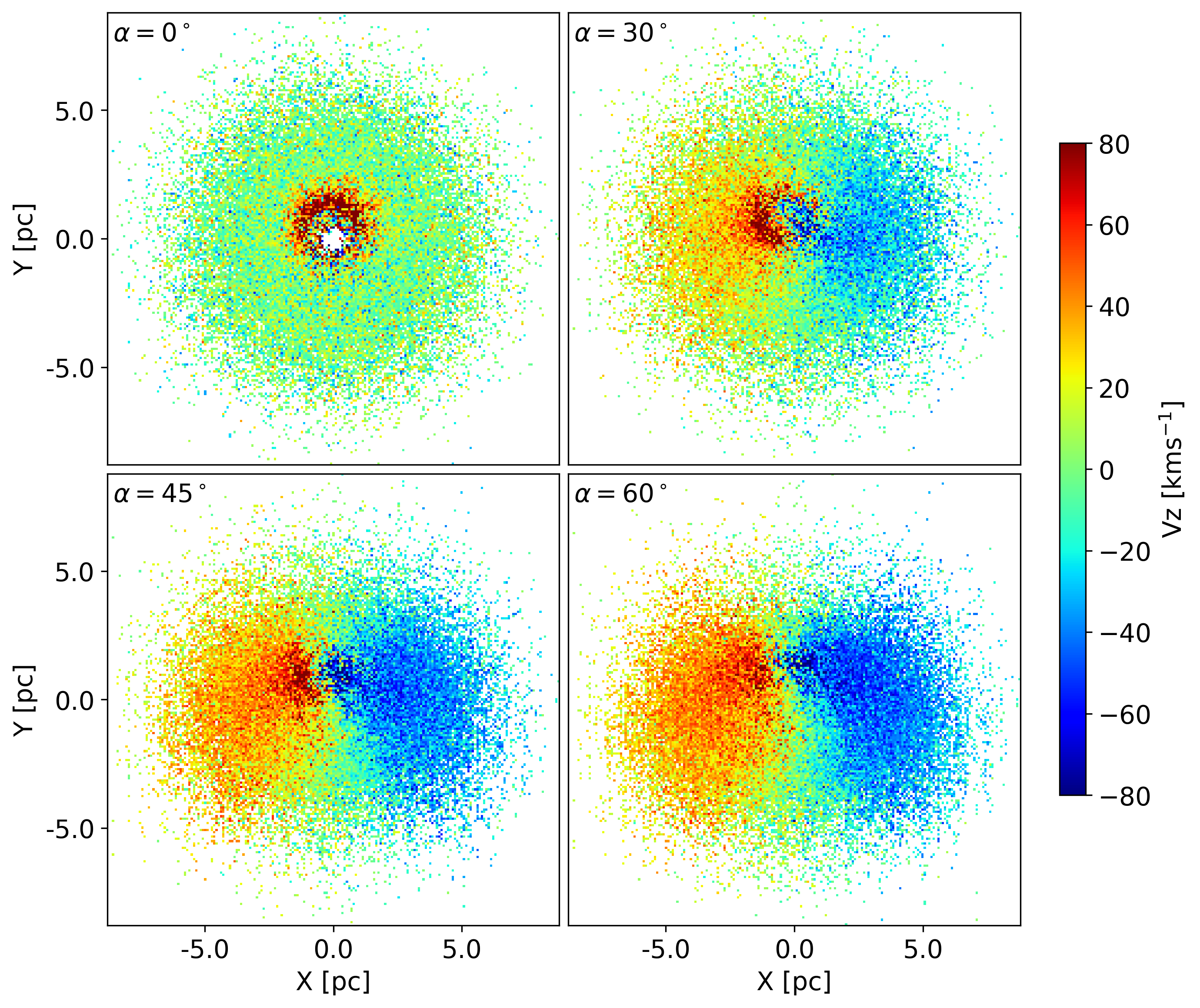}
\caption{ Same as Fig.~\ref{fig:XY90} but $\varepsilon = 0.01$.}
\label{fig:XY90-eps0.01}
\end{figure*} 

\begin{figure*}
\centering
\includegraphics[width = 85mm]{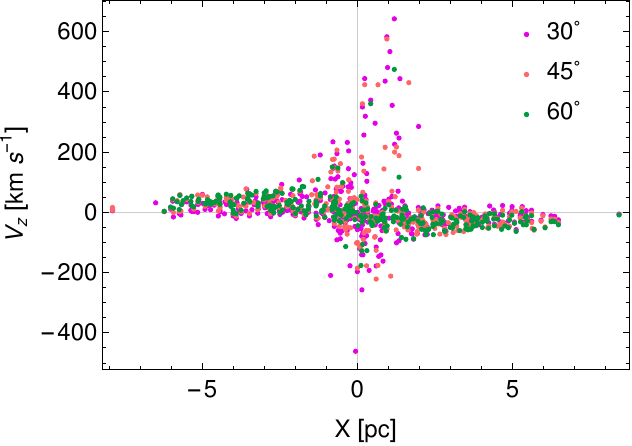}
\includegraphics[width = 85mm]{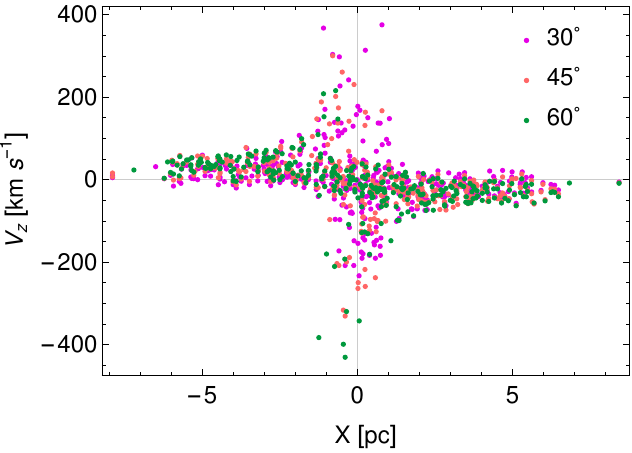}
\caption{Velocity distribution of the clouds in the torus throat for different orientation angles relative to the observer: \mbox{$\alpha = 30^\circ, 45^\circ, 60^\circ$} and relative cloud size $\varepsilon_\text{cl}=0.025$. Parameters of the wind  are: \mbox{$k=20$}, $\gamma=30^\circ$, and azimuthal angle $\phi=0^\circ$ ({\it left}), $\phi=90^\circ$ ({\it right}).}
\label{fig:vel}
\end{figure*} 

\begin{figure}
\centering
\includegraphics[width = 80mm]{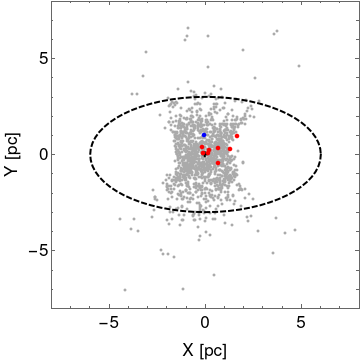}
\caption{Projection of the cloud distribution in the torus throat for $\alpha = 45^\circ$ (gray points). The red and blue points  correspond to the high velocity clouds that can be seen by the observer: \mbox{$V_z > 250$\kms} (red),
\mbox{$V_z < -250$\kms} (blue).  The dashed black ellipse show the conditional boundaries of ALMA emitting region.  The parameters of the wind correspond to the Fig.~\ref{fig:XY0}.}
\label{fig:2vel}
\end{figure} 

To produce velocity maps that correspond to the observational ones, we need to take into account both torus orientation and clouds obscuration, as done in \citep{2021MNRAS.503.1459B}. We rotate the torus around the $x$-axis by an angle $\alpha$ from its original face-on position, and the cloud coordinates are transformed according to:
\begin{align}\label{eqcr1}
X&= x;   \nonumber \\ 
Y&= y \cos\alpha - z\sin\alpha;\\
Z&= y \sin\alpha + z\cos\alpha.  \nonumber
\end{align}
The velocity components are
\begin{equation}\label{eqcr2}
{\bf V}=(V_x, V_y, V_z) = \left(\frac{dX}{dt}, \frac{dY}{dt}, \frac{dZ}{dt}\right). 
\end{equation}
In this case, $\alpha=0^\circ$ corresponds to face-on orientation and $\alpha=90^\circ$ to edge-on view; the picture plane is $XY$ and $Z$ coincides with the line of sight.
 
After rotation, in order to account for cloud-cloud obscuration we apply ray tracing from an observer. For that, we divide the $XY$ plane into $400 \times 400$ cells; then  $100 \times 100$ beams are launched in each cell. We account for the obscuring nature of clouds by stopping a beam when it meets a cloud. When this happens, we record the stopping cloud velocity $V_z$. In each cell, we find the mean value and dispersion of $V_z$, which corresponds to the way observational maps are obtained from spectral data. The relative size of each cloud is a parameter; we consider two values, $\varepsilon_{\text{cl}} = 0.025, 0.01$.

Velocity maps for the azimuthal angle $\phi=0^\circ$ and clouds with the larger of the two values of the relative size, $\varepsilon_{\text{cl}} = 0.025$, are presented in Fig.~\ref{fig:XY0}.  Counter-rotation appears in almost all the projections, seen as redshifted clouds on the field of the blueshifted ones. The effect is less pronounced for the azimuthal angle $\phi=90^\circ$ (Fig.~\ref{fig:XY90}) because in this case the wind axis is oriented in such a way that in projection it coincides with the torus symmetry axis (Fig. \ref{fig:windscheme}). The overall influence of the wind and the differences between these two cases are better seen for smaller clouds, $\varepsilon_{\text{cl}} = 0.01$  (Fig.~\ref{fig:XY0-eps0.01} and \ref{fig:XY90-eps0.01}). The reason is that smaller clouds allow us to look deeper into the torus throat and, as a result, to see more clearly the influence of the wind.

These simulated maps mimic the ALMA observational maps, but show only the low velocity component. Fig.~\ref{fig:vel} presents the full velocity gradient  in the torus throat. It is seen that the clouds can have high velocities in the region near to bicones where the condition of collisional pumping due to interaction with the wind can be realized. As the result, the maser emission can form in such clouds. We suggest that this high velocity clouds are the sources of maser emission.

At this point, we can reproduce the shape of the torus throat for the wind parameters corresponding to Fig.~\ref{fig:hollownwind3D} ({\it top}). An X-shaped distribution appears for angles $\alpha>45^\circ$ relative to an observer. Here, we show the case for $\alpha = 45^\circ$  (Fig.~\ref{fig:2vel}) that corresponds to the torus orientation in Type 2 AGNs, among which is NGC~1068. Thus, the X-shaped structure discovered in ALMA 256~GHz continuum \citep{Impellizzeri2019} can be related to the emission of the torus throat observed due to interactions between the clouds and the wind. The red and blue spots correspond to the clouds with high velocities which are not obscured by others clouds and whose emission is able to cross the torus body. There is an asymmetric red/blue cloud distribution such that the number of red clouds is much larger than the number of blue clouds (Fig.~\ref{fig:2vel}). This example may provide an explanation that the asymmetric distribution of maser spots in the torus NGC~1068 presented in \citep{2004ApJ...613..794G} may be a result of asymmetric wind and orientation effects.

\section{Discussion and Conclusions}
\label{sec:conclusions}

We considered $N$-body simulation of a clumpy torus in the gravitational field of a SMBH with radiation pressure force (wind) acting on the clouds in the torus throat.
The results indicates that a counter-rotation can appear if the wind has an asymmetrical structure. 
To construct model velocity maps, we take into account effects of obscuration by  applying ray-tracing from an observer and treating clouds as optically thick.
The main results are the following:
\begin{itemize}
\item the model maps demonstrate an apparent counter-rotation on the general red-blue rotation of the torus if we apply a wind with an axis tilted relative to the torus symmetry axis;
\item the distribution of the clouds in the torus throat has an X-shape for the orientation near to edge-on relative to an observer;
\item there is an asymmetry in the distribution of the clouds with high velocities in the torus throat which can be seen by an observer; 
\item the presence of the asymmetrical wind together with orientation effects and obscuring conditions can account for the observed counter-rotation in the torus of NGC~1068.
\end{itemize}

The result of our simulation shows that the wind asymmetry can mimic counter-rotation found by ALMA in the torus of NGC~1068 \citep{2019A&A...632A..61G}.  This apparent counter-rotation can give us additional information about the wind structure. The wind axis coincides with the angular momentum vector of the accretion disk and the inclination of the wind axis can tell us about the inclination (or precession) of the accretion disk relative to equatorial plane of an AGN. This can be a result of misalignment between the SMBH spin axis and the disk which leads to the complicated geometry of the disk including its titling and formation of inner rings \citep{N2015}. Anisotropy of the wind can also appear due to presence of warp accretion disk. New multi-band mid-IR images of the dust sublimation radius in the torus of NGC~1068 made by \citep{Gamez2022} show that distribution of the warm and hot dust reveals  two main axes. It may indicate the role of the black hole spin in the complicated dynamics in such a system. 

Using our semi-analytical model of the asymmetrical radial wind, presented by radiation pressure, we show the possibility to reproduce an apparent counter-rotation effect. In general, the wind can be even more complicated, for example, with the presence of a swirl in hydrodynamical consideration of wind. Other mechanisms can give additional effects, for example, the influence of a binary SMBH or external accretion.  The torus plays a role of a reservoir, feeding the accretion disk and providing the high luminosity of AGN. In this case the external accretion is important to supply torus with matter. \cite{2019A&A...632A..61G} estimated that mass of the outflowing gas inside the torus of NGC 1068 is comparable with that of the molecular gas in streams that supply material to the torus. This demonstrates the complexity of the system which requires the including an external accretion.  If the direction of the external accreted matter does not coincide with orbital direction of the torus it can be an additional reason for counter-rotation. We plan to check these scenarios in our future simulations.

\section*{Acknowledgements}
\addcontentsline{toc}{section}{Acknowledgements}
EB and VA are very grateful to all Italian colleagues from different  Astronomical Observatories of Italian National Institute for Astrophysics (INAF) for the help during our evacuation from Kharkiv and for the following support in Italy. It provided us with a place to work (EB in Capodimonte Astronomical Observatory and VA in Turin Astronomical Observatory) and allowed us to finish this work. We also grateful the colleagues of Astronomick{\'e} observat{\'o}rium na Kolonickom sedle (Slovakia) for the help and hospitality.
The work of EB, PB, VA and MI  was supported under the special program of the NRF of Ukraine `Leading and Young Scientists Research Support' -- `Astrophysical Relativistic Galactic Objects (ARGO): life cycle of active nucleus', No. 2020.02/0346.
PB and MI express their great thanks for the hospitality of the Nicolaus Copernicus Astronomical Centre of Polish Academy of Sciences where some part of the work was done.
The work of PB and MI was supported by the Volkswagen Foundation under the grant No.~97778. The work of PB was also supported by the Volkswagen Foundation under the special stipend No.~9B870 (2022).
PB and MI also acknowledges support by the National Academy of Sciences of Ukraine under the Main Astronomical Observatory GPU computing cluster project No.~13.2021.MM.
The work of PB and MI was supported by Ministry of Education and Science of Ukraine under the France - Ukraine collaborative grant M63-17.05.2022 (DNIPRO). 
PB and MI acknowledges the support from the Science Committee of the Ministry of Education and Science of the Republic of Kazakhstan (Grant No.~AP14870501).

\section*{Data availability}
\addcontentsline{toc}{section}{Data availability}
Simulation data and codes used in this paper can be made available upon request by emailing the corresponding author.

%%%%%%%%%%%%%%%%%%%% REFERENCES %%%%%%%%%%%%%%%%%%

\bibliographystyle{mnras}
\bibliography{torus} % if your bibtex file is called example.bib

\bsp	% typesetting comment
\label{lastpage}
\end{document}